# Cloud Deployment of Large-Scale Electromagnetic Transient Simulation – Discovery and Experiences

Xiaochuan Luo, Jason Ploof,
Xinghao Fang, Qiang "Frankie" Zhang
ISO New England Inc.
(xluo, jploof, xfang, qzhang@iso-ne.com)

Song Zhang
Energy & Utilities, Amazon Web Services
songaws@amazon.com

*Abstract*—Electromagnetic Transient (EMT) simulation starts to play a critical role in modern power system planning and operations due to large penetration of inverter based resources (IBRs). The EMT studies are computationally intensive due to very small simulation time step and complex modeling of the protection and control of IBRs. It has been challenging for the traditional on-premises computing infrastructure to meet the ever-increasing computing needs of large-scale EMT studies. This paper shares experience of ISO New England (ISO-NE) on a pilot deployment of EMT simulation in a public cloud using Amazon Web Services. The platform can successfully meet the large-scale EMT simulation computation needs in a cost-effective way while meeting cyber security and data privacy requirements.

*Index Terms*-- Electromagnetic Transient, Amazon Web Services (AWS), Cloud computing

## I. Introduction

New England is rapidly transitioning from traditional sources of energy to renewables. As shown in Figure 1, out of over 35 GW in ISO-NE's generation interconnection queue as of June 2023, 97% of them are IBRs consisting of wind, battery, and solar plants. In addition, behind-the-meter PV installations across the region will be grown to around 6 GW by the end of 2023, and are projected to grow to 12 GW by 2032, as shown in Figure 2 [1]. Vineyard Wind, the largest offshore wind plant in the US with a total capacity of 800 MW, is expected to be on-line in New England by the end of 2023.

Typically, traditional transient stability (TS) analysis has been used to study the dynamic behavior of the power system. It is a positive-sequence root mean square (RMS) based modeling and simulation approach. This approach has proven effective for studying power systems dominated by conventional synchronous generators. However, it cannot adequately capture some of the unique performance characteristics of IBRs. EMT modeling and simulation has started to become more popular in recent years to study IBR control interactions, sub-synchronous oscillations, weak grid instability, and fault ride-through performance, etc. These EMT simulations are very computationally demanding due to the following reasons: (1) extremely small time step of around 10 microseconds, in order to capture fast electro-magnetic transients, switching actions in the kHz-range of power electronics devices such as Insulated-gate Bipolar Transistor (IGBT), protection actions, and dynamics of fast controls such as Phase-locked Loop (PLL); (2) sophisticated EMT models for power system components with extensive details so as to accurately describe fast or non-linear system behavior, such as inverter control, transformer magnetic saturation, wave propagation along overhead transmission lines and underground cables, etc.; (3) core algorithm targeting solving differential equations for each circuit branch in finding instantaneous three-phase electrical quantities.

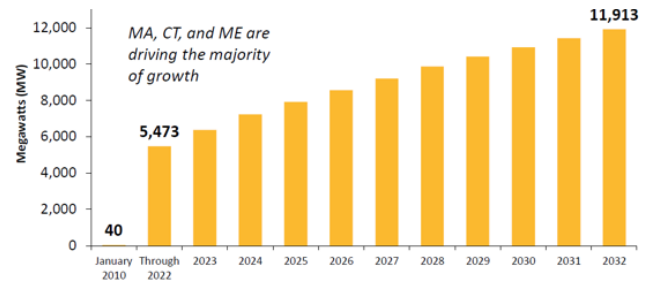

Figure 2. ISO-NE's forecast of PV growth through 2032

A typical EMT simulation run normally takes much more time (by a few orders of magnitude) than a TS simulation, hours vs. minutes. Therefore, EMT simulations usually only model a portion of the power system in the electrical vicinity of the study area, with the rest of the system represented by a static equivalent. This approach reduces the computational demands at the cost of the observability of wide-area dynamic phenomena. Even so, sometimes the EMT models could inevitably become very large, which further prolongs the simulation time. For example, the Australian Energy Market Operator (AEMO)'s EMT case has 3700 buses and 140 detailed IBR models; ISO-NE's Western Massachusetts (a sub-region) DER cluster case alone has 485 buses and 94 detailed models [2].

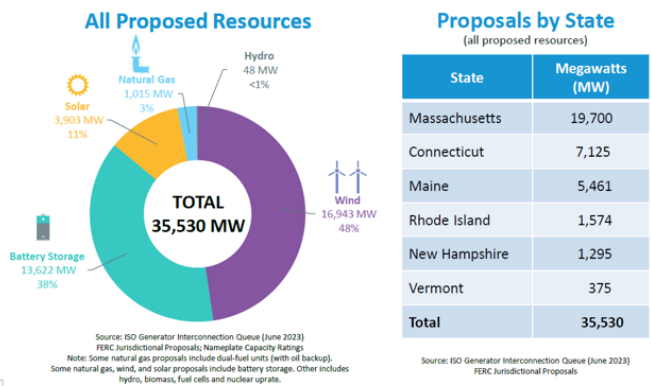

Figure 1. ISO-NE interconnection queue as of June 2023

The computational challenges facing EMT simulation is not new since traditional power system studies could also take a long time if we run many scenarios and contingencies. ISO-NE has been running a high performance computing (HPC) - Cloudfuzion platform on AWS for large-scale planning studies for more than seven years [3]. Power system software deployed in this platform includes Siemens PSS/E for transient stability studies, GE MARS for resource adequacy analysis, and PowerGem TARA for large-scale N-1-1 contingency analysis. This cloud based HPC platform has tremendously increased the efficiency of our planning studies in a cost-effective way. By providing users with the capability of provisioning hundreds of servers under a flexible "pay-as-you-go" pricing scheme, cloud computing opens up a new opportunity for engineers to conduct more in-depth power system studies that cannot be effectively done by on-premises servers due to internal IT infrastructure limitations and time constraints. The successful use of cloud computing prompted us with the idea of using it for the EMT simulation.

In this paper, we discuss in detail how ISO-NE implemented a cloud based EMT simulation platform using Amazon Web Services (AWS) AppStream 2.0 service, which is a pioneering implementation and to our knowledge the first in the power industry to deploy both PSCAD and E-Tran for parallel EMT simulation in the cloud. The paper is organized as follows. In Section II, the design of the ISO-NE's cloud based EMT simulation platform is presented. An EMT case study for performance comparison between the cloud and on-premises environment is discussed in Section III. Section IV discusses the security aspects of the implementation, and Section V provides further discussions and conclusions.

## II. Design of Cloud Based EMT Simulation

According to US National Institute of Standards and Technology (NIST), cloud computing is defined as "a model for enabling ubiquitous, convenient, on-demand network access to a shared pool of configurable computing resources (e.g., networks, servers, storage, applications, and services) that can be rapidly provisioned and released with minimal management effort or service provider interaction" [4]. In accordance with ownership, cloud computing comes in three forms: public, private, and hybrid clouds. From the service model perspective, cloud computing can be primarily offered as Infrastructure as a Service (IaaS), Platform as a Service (PaaS), and Software as a Service (SaaS). The operational efficiency, resilience, and sustainability of cloud computing makes it an attractive IT solution for many organizations, including many utilities and solution providers in the power industry [5].

Considering the large penetration of IBRs and the challenges with EMT modeling and simulations, ISO-NE had an EMT Initiative on the corporate scorecard in 2022. The scorecard project included three parts: EMT model management, EMT-TS hybrid simulation, and cloud-based EMT simulation. The main objectives of the cloud-based EMT simulation part are to demonstrate the feasibility of large-scale EMT simulation in the cloud, compare its performance with our on-premises computing environment, and identify any constraints and limitations on EMT cloud deployment. Based on the experience gained through the pilot deployment, a decision will be made in the future about production level implementation for our planning and operations studies.

### A. Requirements and Choice of AWS Services

There are multiple established Cloud Service Providers (CSPs) on the market. The top three leaders are Amazon Web Services (AWS), Microsoft Azure and Google Cloud. As ISO-NE has already had an enterprise-ready, multi-account AWS environment and a production-level high-performance computing (HPC) workload hosted on AWS, which has been running reliably and securely for years, we decided to continue using AWS for the pilot EMT simulation cloud deployment.

The production software of EMT simulation at the ISO-NE includes PSCAD and E-tran. Due to the constantly high CPU utilization rate and excess TCP communication overhead during the PSCAD parallel simulations, it is deemed that a dedicated tenancy on the running instance would be preferable. In other words, each user should occupy a Virtual Machine (VM) exclusively. If more users need to conduct a PSCAD study, additional VM instances should be launched and allocated to meet the requests. In addition, it is desirable to provide secure desktop streaming to users, so they can play on their own EMT simulation workbench via a browser with ease for tasks such as modeling, parameter configuration, simulation start/stop, etc.

Technically there are two ways for a user to access a cloud instance's desktop: through Remote Desktop Protocol (RDP) and by means of streaming protocol based on Transport Layer Security (TLS). The latter is usually provisioned as Virtual Desktop Infrastructure (VDI) and Desktop-as-a-Service (DaaS) in the cloud. On AWS, the DaaS service can be further divided into two subcategories: Desktop View End User Computing (e.g., AWS WorkSpaces and AWS Workspaces Web) and Application View End User Computing (e.g., AWS AppStream 2.0 [6]). At the time of the project, ISO-NE didn't have Site-to-Site VPN or other private connectivity between the corporate data center and AWS cloud. RDP to a cloud instance that sits outside ISO's corporate network without VPN is not allowed per IT and cybersecurity requirements. For this reason, AppStream 2.0 was chosen as the hosting service for the EMT simulation platform because it offers a simple and secure solution to application-centric desktop delivery. It is safer than AWS WorkSpaces due to the nature of non-persistent virtualization [7]. AppStream 2.0 offers more choices of instance types characterized by CPU, memory and GPU optimized to reduce costs and the option for launching the instance inside an Amazon Virtual Private Cloud (VPC). This feature is crucial to enable Bring-Your-Own-License (BYOL) for commercial software in the AWS environment because a dedicated licensing host needs to be spun up in the same VPC to communicate with the application sessions started by AppStream 2.0. Due to the ephemeral nature of AppStream 2.0, the licensing host was started in Amazon Elastic Compute Cloud (also known as EC2) in this use case.

B. *AWS Environment Setup*

Figure 3 is the high-level architecture diagram of the AWS environment setup that includes several major components:

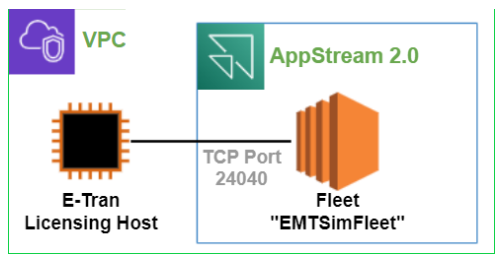

Figure 3. High-level architecture diagram

1) *Virtual Private Cloud (VPC): A* virtual network closely resembles a traditional network that we would operate in our own data center, with the benefits of using the scalable infrastructure of AWS.

2) *E-Tran Licensing Manager:* Software installed on an on-demand EC2 instance (t2.small) with port 24040 enabled. After license files (Etran6.lic, EtranPSCAD.lic, ETranPSSE.lic) were imported, the License Manager program started to listen on port 24040 for licensing.

3) *AppStream Image:* The following software were installed on Appstream 2.0 image using Image Builders:

- PSCAD v4.6.3 with Certificate License, which is through ISO-NE's PSCAD MyCenter Web portal account
- Microsoft Visual Studio 2010 Express for model editing compatible with the corresponding PSCAD version
- E-Tran Plus for PSCAD

Note that only pre-compiled PSCAD simulations were run for simplicity, in order to avoid installing other required compilers to build the case. The focus here is on the performance of the simulation in the cloud.

4) AppStream Fleet and Stack: Once an AppStream 2.0 Image was created, a fleet named "EMTSimFleet" was created using the image and selected Instance Type for the end users to connect to the streaming desktop. A stack named "EMTSimEnvStack" was also created to control the user access and permission to the streaming fleet. After a user was added to the Stack's User Pool, he/she would receive an email invitation with AppStream login URL, user name and temporary password.

C. *User Access and Initial Testing*

When a user logs into an AppStream instance via a web browser using an https URL link, a streaming instance is spun up from a prebuilt AppStream Image, which has all the needed software (PSCAD and E-Tran in our case) installed and configured. Each user will exclusively occupy an instance. If another user connects to the service simultaneously, AppStream automatically scale out the fleet size to match the supply of available instances to user demand. The GPU, CPU, memory, storage, and networking capacity of each instance is specified in the Fleet configuration. Once the number of concurrent users exceed the fleet maximum capacity limit, the new connection request will be rejected until any connected user's session timeout. It's easy to plan the fleet capacity for the EMT study based on the expected size of workforce. In case of any program version changes or additional required programs to be installed, we only need to update the AppStream Image.

The first issue that was quickly discovered and resolved was that the CPU allocation to the instance did not suffice what was needed. The compute-optimized instance type *stream.compute.4xlarge* was first selected because of its high performance processors. However, the instance has only 16 virtual processors and 30GB memory, which proved to be too slow. Later, two other types of instances with better performance were chosen: *stream.compute.8xlarge,* which is compute optimized with 32 virtual processors and 60GB memory, and *stream.memory.z1d.12xlarge,* which is memory optimized with 48 virtual processors and 384GB memory. This is one of the main advantages the cloud brings over traditional computing infrastructure; the selected instance types for the workload can be easily upgraded if necessary (sometimes at a reasonable cost difference).

The second issue we discovered was that the simulation frequently crashed in the middle if PSCAD output was enabled. The output files are essential since they contain results from the simulation and are used for plotting and engineering analysis. Upon further investigation, we found that the read/write speed of the default data storage (Home Folder) for AppStream was Amazon S3, which is not suitable for workload requiring high I/O per second and large throughput. It was too slow for the amount of data being produced by the EMT simulation. As the simulation buffered output data into batches and then wrote them to storage, the write speed limitations caused an increasing backlog and started to fill the cached memory. Eventually the backlog made the cached memory overflow and caused the simulation to crash.

After researching other storage services on AWS, an alternate file management system, FSx to be specific, was adopted and set up to overcome the disadvantages of S3. FSx provides a Microsoft Windows file server backed up by a fully native Windows file system. This file management system allows the administrator to set a storage capacity (GB) and the throughput capacity (MB/s) that the file system runs on. The throughput was our main focus, since this was our limiting factor as mentioned before. An FSx for Windows file system with SSD storage type with 32 MB/s throughput capacity was provisioned and linked to the AppStream fleet. With this change, the simulations were able to finish their runs and the output files were produced properly, even when a smaller plotting time-step was used that would increase the throughput.

## III. BENCHMARK TESTING AND PERFORMANCE

We have tested and compared the EMT simulation performance across four platforms. The specifications of the platforms are summarized in Table 1.

Table 1. Specification of tested hardware platforms

|  | *Premium Mini Workstation* | *On-premises HPC Server* | *compute.8 xlarge* | *z1d.12xlarge* |
|---|---|---|---|---|
| CPU model | Intel i7-10700 | Intel Xeon Platinum 8176 server (dual socket) | Intel Xeon E5-2666 (dual socket) | Custom Intel Xeon Platinum 8151(dual socket) |
| CPU cores/threads | 8/16 | 28/56 (56/112) | 8/16 (16/32) | 12/24 (24/48) |
| CPU base/boost frequency (Hz) | 2.9/4.8 | 2.1/3.8 | 2.9/3.5 | 3.4/4.0 |
| First release date | Q2 2020 | Q2 2019 | Q1 2015 | Q2 2019 |
| L1/L2/L3 Cache | 256KB/1MB/16MB | 1.5MB/28MB/38.5MB | 640KB/2.5 MB/25MB | Not Disclosed* |

*No public info about the custom processor disclosed by Amazon

The premium mini-workstation and on-premises HPC server are on-premises at the ISO-NE, while *stream.compute.8xlarge* and *stream.memory.z1d.12xlarge* are AWS AppStream instances. The testing suite was PSCAD simulation cases from System Impact Study of a wind plant in ISO-NE region, including subcases with 1, 2, 4, 6, 12, and 24 inverters, all of which were parallelized in the simulation. These cases were run at a time step of 10μs and a plot time step of 50μs, and all the simulations were pre-compiled and run with a GUI-less setup.

Figure 4 is a comparison of EMT simulation runtime among the four platforms. It can be seen that when the number of inverter models was less than or equal to six, the performance (in terms of simulation speed) of the four environments was comparable with *z1d.12xlarge* as the fastest. This is because when the number of inverter models is less than the number of CPU cores, each inverter can be allocated to one core for parallel simulation. In such cases, the CPU frequency and cache memory are the determining factors for simulation speed.

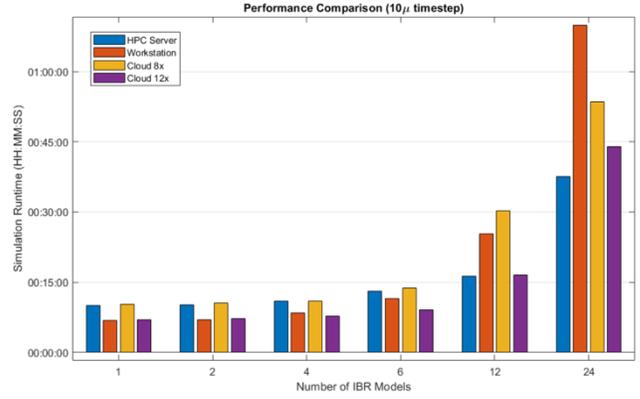

Figure 4. Simulation run time comparison across four platforms

When the number of inverter models increased to 12, the simulation on the mini workstation started to increase noticeably, while the time spent on the on-premises HPC server and AWS *z1d.12xlarge* instance did not have an obvious rise. The mini workstation slightly outperformed the AWS compute.8xlarge instance despite possessing on fewer cores because its processors has much newer architecture based on the CPU release date. When the number of inverter models further increased to 24, the HPC server and two types of cloud instances far led the workstation because they had enough cores for a "1-to-1" inverter allocation, as opposed to the mini workstation having to oversubscribe the subcase simulation across the available cores, which led to a notable slowdown of the simulation. The on-premises HPC server with 56 cores slightly outperform the *stream.memory.z1d.12xlarge* instance with 24 cores because the latter has all cores fully occupied with less room for other tasks such as I/O burst and TCP port communication overhead.

It should be noted that z1d.12xlarge is not the best option on AWS for CPU-bound tasks such as EMT simulation. Due to the time constraint, this work only included the experiment results that were collected so far. However, based on our prior experience, we can reasonably assume that with more flexible choices of VM instances for HPC purpose such as the latest EC2 Hpc7a instances, the cloud will outperform the on-premises HPC server as we continue to increase the number of inverters due to the scalability and always up-to-date infrastructure on the cloud.

## IV. SECURITY SCHEMES

Security is the top concern for cloud users, especially when it comes to a regulated industry such as power industry. Most CSPs use Shared Responsibility Model (SRM) for security and compliance, which means that CSP is responsible for the security "of the cloud" including protecting the infrastructure that runs all of the services, while the customer is responsible for securing what is "in the cloud," such as their data, network and security configuration as well as task management. The responsibility also differs depending on the cloud deployment model. Due to the complexity and many misconceptions of cloud security, the power industry has been conservative in adopting the cloud technology. In this work, we have employed a comprehensive security scheme to protect the cloud-based EMT simulation platform.

1) *VPC deployment*: Amazon VPC lets the user provision a logically isolated section of the AWS. The user has complete control over the virtual networking environment, including selection of IP address range, creation of subnets, and configuration of route tables and network gateways. We defined specific VPC and subnet for EMT AppStream Fleet with network access control (ACL) so that the IP traffic can be controlled and monitored at the network interface.

2) *Security group*: A security group acts as a virtual firewall to control the inbound and outbound traffic for cloud instances. We created a specific security group for the EMT AppStream Fleet. It defines inbound and outbound traffic that are only allowed from the IP addresses owned by ISO-NE, through dedicated ports, and using specific protocols. All other communication paths are completely blocked.

3) *Data encryption*: To prevent sensitive data that contains critical energy infrastructure information (CEII) from being intercepted while transmitting between corporate network and the cloud, the traffic is encrypted based on HTTPS protocol with a dedicated SSL certificate. As for the simulation data at rest in the cloud, they are encrypted in the FSx for Windows file system and S3 bucket using one of the strongest industry standard encryption algorithms (i.e., AES-256) to ensure the data privacy.

4) *Identity and Access Management (IAM):* AWS's IAM enables security control access to AWS services and resources. ISO-NE's IT administrators used Role-based Access Control (RBAC) to permit access to AWS resources. In addition, a two-factor authentication was established for every user to add an extra layer of protection. For engineers who run EMT simulation via AppStream, the administrator created Appstream User Pool with each user's unique ISO-NE email address and password to access the AppStream.

5) *PSCAD Software:* PSCAD Certificate License was used in our cloud-based EMT simulation. The Certificate License requires the user to log in and check out the license from PSCAD License Management Portal, i.e. MyCenter. When a user finishes the session, he/she must release the license back to MyCenter.

Overall, these security schemes provide comprehensive protection over the cloud resources, confidential CEII information, and power system application software.

## V. Discussion and conclusion

In early 2023, ISO-NE finished its corporate AWS Foundation project. One of the key accomplishments is establishing two dedicated connections from AWS to the ISO-NE on-premises network, which is known as Direct Connect. With the connection, AWS can be treated as an extended data center of ISO-NE with high bandwidth, enhanced security, and more consistent network experience than internet-based connections. Next, we plan to redesign our pilot implementation with a few potential improvements as follows:

1) *Remote Desktop Protocol (RDP):* with Direct Connect, the outbound traffic from any ISO-NE host to AWS will no longer tranverse public internet. We do not have to use AppStream for desktop streaming; instead, RDP becomes an viable option so we can unlock EC2 service as the hosting environment because it has many more instance type options to rightsize the computing resources compared with AppStream.

2) *PSCAD and E-Tran License:* with Direct Connect, we can use our on-premises E-Tran and PSCAD network license server to directly license the cloud instances. There is no need to set up a dedicated lincese server on AWS, and no need to use the PSCAD Certificate License. One issue we encountered using the Certificate License was that the engineer forgot to return the license when they disconnected with the AppStream session or the session timed out. This led to accidentally holding onto a PSCAD license in the cloud, while other users were unable to accquire a license because only a limited number of Certificate Licenses were available.

3) *Intel Fortran Compiler:* In this pilot implementation, we were running GUI-less PSCAD simulations because Intel Fortran was not installed in the cloud per our IT policy and user agreement. GUI-based PSCAD simulation is not desirable in an HPC environment since it occupies an extra core that can be allocated to an inverter model instead and users don't need to look into the results until the whole simulation completes. However, if we plan to have a full transition to cloud-based PSCAD simulation, the Intel Fortran Compiler is necessary to build the cases and initialize files, and it has to be installed together with PSCAD on AWS to streamline the workflow. With the completion of ISO-NE's AWS Foundation Project, IT has established internal processes and procedures to create cloud golden image which we can utilize to deploy Intel Fortran Compiler on AWS.

Overall, our pioneering implementation of cloud-based EMT simulation was a success. It has clearly demonstrated the feasibility of large-scale EMT simulation in the cloud, the benefits of using cloud technology with sufficient security mechanism to protect confidential CEII data, and some future improvement for the architecture design. Cloud technology has been evolving fast in recent years, while the power industry has been slow to adopt cloud-based solutions. For non-mission critical workloads which are not subject to compliance and regulatory requirements such as NERC CIP, the industry could fully take advantage of the benefits that cloud technology brings to accelerate the grid transformation in decarbonization, digitalization and decentralization.